\documentclass[twocolumn,aps,prl,reprint,groupedaddress]{revtex4}
\usepackage{graphicx}
\usepackage{color}
\usepackage{amsmath}

\newcommand{\be}{\begin{equation}}
\newcommand{\ee}{\end{equation}}

\newcommand{\bea}{\begin{eqnarray}}
\newcommand{\eea}{\end{eqnarray}}

\newcommand{\p}{\partial}

\newcommand{\lp}{\left(}
\newcommand{\rp}{\right)}

\renewcommand{\vec}[1]{{\bf #1}}
\renewcommand{\hat}[1]{{\widehat #1}}

\renewcommand{\tensor}[1]{\underline{\mathbf #1}}
\newcommand{\unit}{\vec 1}

\begin{document}
\title{Hall Drag and Magnetodrag in Graphene} 

\author{Justin C. W. Song$^{1,2}$}
\author{Leonid S. Levitov$^1$}
\affiliation{$^1$ Department of Physics, Massachusetts Institute of Technology, Cambridge, Massachusetts 02139, USA}
\affiliation{$^2$ School of Engineering and Applied Sciences, Harvard University, Cambridge, Massachusetts 02138, USA}





\begin{abstract}
Massless Dirac fermions in graphene at charge neutrality form a strongly interacting system in which 
both charged and neutral (energy) modes play an important role. These modes
are essentially decoupled in the absence of a magnetic field, but become strongly coupled when a field is applied. 
We show that these ideas explain the recently observed giant magnetodrag, arising in classically weak fields when electron density 
is tuned near charge neutrality. We predict strong Hall drag in this regime, which is in stark departure from the weak coupling regime, where theory predicts the absence of Hall drag.
Energy-driven magnetodrag and Hall drag arise in a wide temperature range and at weak magnetic fields, and feature an unusually strong dependence on field and carrier density.
\end{abstract}


\maketitle
Graphene near charge neutrality (CN) hosts an 
intriguing electron-hole system with unique properties\cite{gonzalez,sheehy,son,vafek2007,kashuba,fritz,mueller,zuev,wei,checkelsky}.
Our understanding of the behavior at CN would greatly benefit 
from introducing ways to couple the novel neutral modes predicted at CN to charge modes which can be easily probed in transport measurements. There is a long history
of employing magnetic field for such a purpose, since transport in
charge-neutral plasmas is ultra-sensitive to the presence of external magnetic fields\cite{PitaevskiiLifshitz}.

A new interesting system in which magnetotransport  at CN can be probed 
are atomically thin
graphene double layer G/hBN/G structures\cite{britnell1,geim}. 
Strong Coulomb coupling between adjacent layers in these systems results in strong 
Coulomb drag,
arising when current applied in one (active) layer induces a voltage in the adjacent (passive) layer\cite{geim,tutuc,tse2007,narozhny2007,hwang2011,peres2011,katsnelson2011,edrag,narozhny2012,magneto-dima}. 
Recent measurements\cite{geim} revealed  
drag resistance that peaks near CN and has  dramatic magnetic field dependence, with the peak value increasing by more than an order of magnitude (and changing sign) upon application of a relatively weak $B$ field. 
Strong magnetic field dependence of drag has been observed previously in other double layer two-dimensional electron gas (2DEG) heterostructures\cite{patel1996,rubel1997,lilly1998}, however these experiments were carried out in the quantum Hall regime, whereas the anomalous magnetodrag found in Ref.\cite{geim} occurs at classically weak fields $B\lesssim 1\,{\rm T}$. 

\begin{figure}[t]
\includegraphics[scale=0.145]{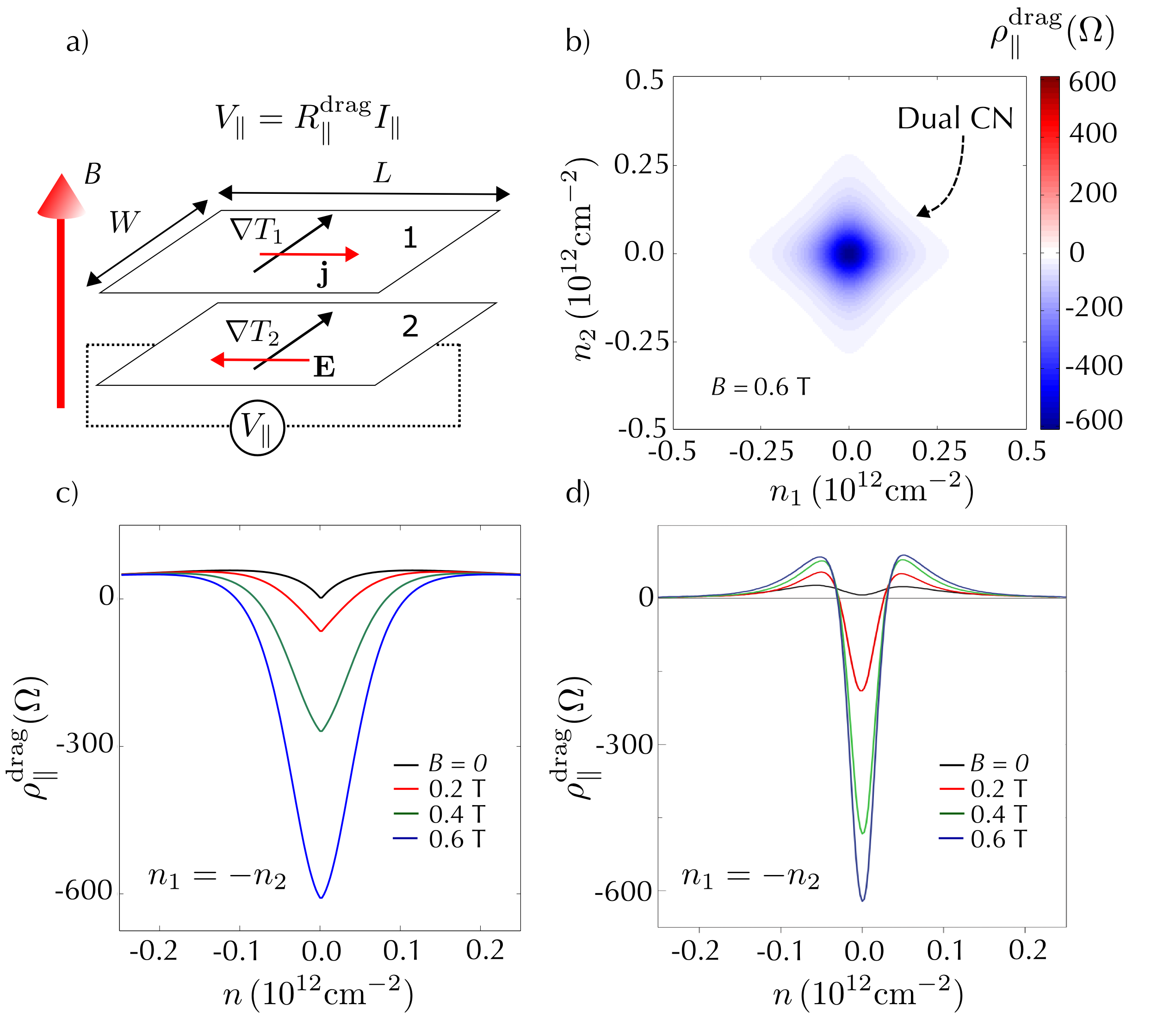} 
\caption{Energy-driven magnetodrag in a double layer graphene heterostructure 
close to CN. (a) Schematic of charge current, temperature gradients,  and electric field in the two layers that give rise to a {\it negative} $\rho_{\parallel}^{\rm drag}$. (b,c) 
Magnetodrag resistivity, $\rho_{\parallel}^{\rm drag}$, obtained from Eqs.(\ref{eq:draguniform}),(\ref{eq:resistivity}).  Parameter values: $B=0.6 \, {\rm T}$,  $n_0 = 10^{11} \, {\rm cm}^{-2}$, $T = 150 \, {\rm K}$, and $\rho_0 =  \frac{h}{3 e^2}$. 
The $B=0$ dependence taken from the model of drag at zero B field \cite{edrag,narozhny2012}.(d) Experimentally measured magnetodrag resistivity in G/hBN/G heterostructures, reproduced from Ref. \cite{geim} for the same $B$ values as in (c). Application of magnetic field leads to a giant negative drag at CN. Note the similarity between data and theoretically predicted drag density dependence, $B$ dependence, and sign. }
\label{fig1}
\vspace{-5mm}
\end{figure}

Here we explain this puzzling behavior
in terms of an energy-driven drag mechanism which involves coupled energy and charge transport\cite{edrag,magneto-dima} (see Fig.\ref{fig1}). Energy transport plays a key role because of 
fast vertical energy transfer due to interlayer Coulomb coupling in G/hBN/G systems\cite{edrag}
and relatively slow electron-lattice cooling\cite{graham2013,betz2013}.
As a result, current applied in one layer can create a spatial temperature gradient for electrons {\it in both layers}, giving rise to thermoelectric drag voltage. The effect peaks at CN, since thermoelectric response is large close to CN\cite{zuev,wei,checkelsky} and diminishes as $1/E_F$ upon doping away from CN\cite{ziman,hwang2009}.
Drag arising from this mechanism depends on thermoelectric response and, unlike the conventional momentum drag mechanism, it is insensitive to the electon-electron interaction strength.

Another interesting effect that 
can be probed in these systems is that of Hall drag. It has long been argued that, at weak coupling, no Hall voltage can arise in the passive layer in the presence of  current in the active layer \cite{kamenev1995,bonsager1996}. 
This is because transferred momentum 
is parallel to velocity, allowing only a longitudinal ``back-current" to develop in the passive layer.
As we shall see, a very different behavior arises at strong coupling, owing to
the long-range energy currents
leading to electron-lattice temperature imbalance. Close to 
CN, the magnitude of the cross-couplings between charge and energy currents becomes large, producing a finite Hall drag,
$
V_H = R_H^{\rm drag} I_\parallel
$. 

\begin{figure}
\includegraphics[scale=0.145]{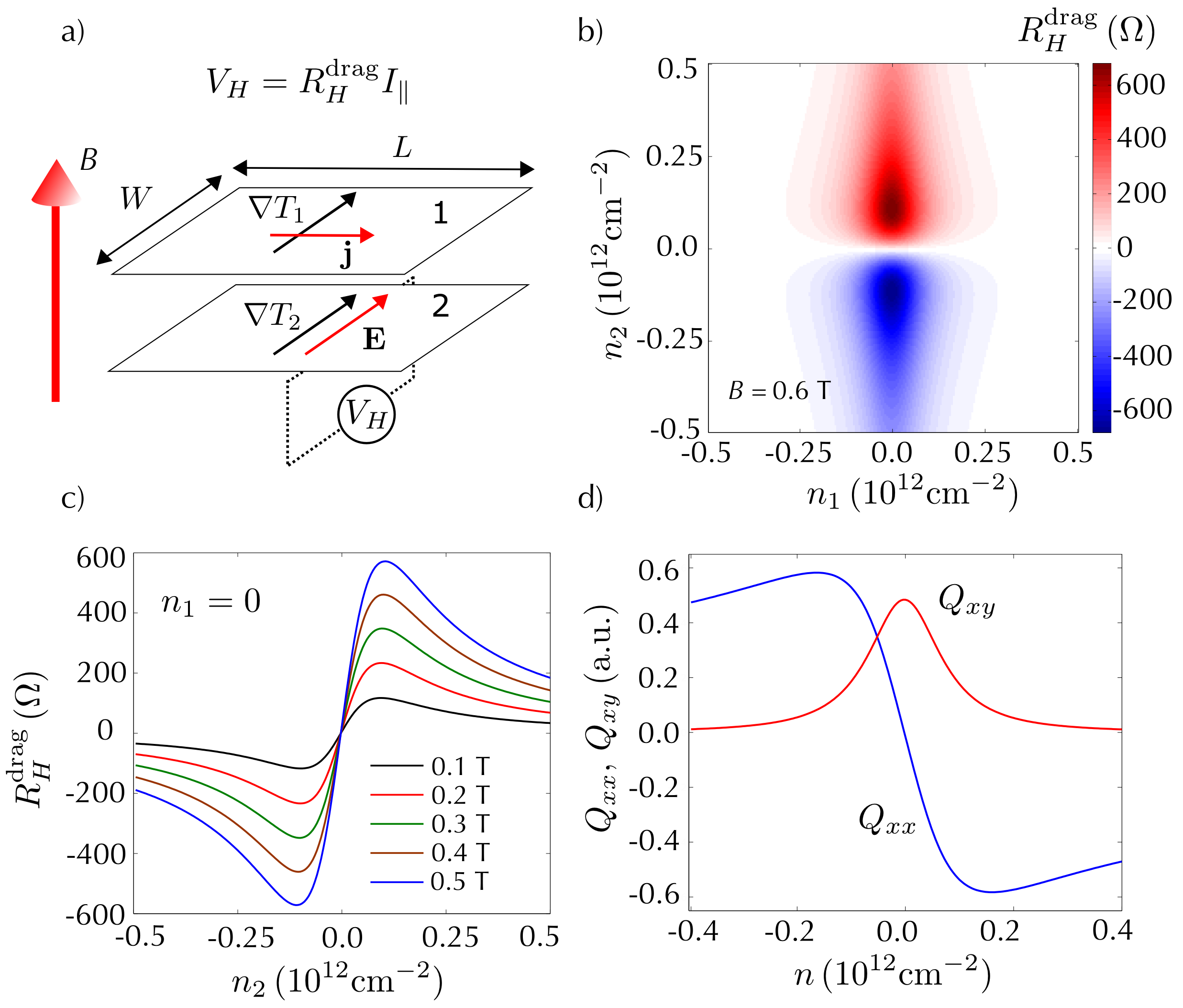}
\caption{
(a) Schematic of charge current, temperature gradients,  and electric field in the two layers of a Hall bar that produces Hall drag. (b,c) Density dependence of Hall drag resistance,  
predicted from Eqs.(\ref{eq:draguniform}),(\ref{eq:resistivity}) for the same parameter values as in Fig.\ref{fig1}.
(d) Density dependence of $Q_{xx}$, $Q_{xy}$, see text.}
\label{fig2}
\vspace{-5mm}
\end{figure}

As we will see, energy currents result in Hall and magnetodrag resistances, $R_H^{\rm drag}$ and $R_\parallel^{\rm drag}$, that are large and peak near CN, see Fig.\ref{fig1} and Fig.\ref{fig2}. 
These large values arise even for classically weak fields $B\sim 0.1 \, {\rm T}$,
exceeding by two orders of magnitude the values found previously in GaAs/AlGaAs 2DEG heterostructures \cite{patel1996,rubel1997,lilly1998} at similar fields. The mechanism based on coupled energy and charge transport predicts large and negative 
drag at CN that matches recent experiments  (see Fig.\ref{fig1}c,d). 
Our mechanism naturally leads to Hall drag because vertical energy transfer between layers does not discriminate between longitudinal and transverse heat currents since temperature profile is a scalar field. This stands in contrast to conventional momentum driven drag, where momentum transfer is parallel to the applied current \cite{kamenev1995,bonsager1996}.

Heat current and an electric field, induced by charge current and temperature gradients, are coupled via the thermoelectric effect altered by the $B$ field,
\be
\vec {j_q} = \tensor{Q} \vec j, \quad \vec{E} = \tensor{Q}\frac{\mathbf{\nabla}T}{T} .
\label{eq:jqande}
\ee
Here $\tensor{Q}$ is a $2\times 2$ matrix, of which 
diagonal components 
describe the Peltier and Thompson effects, and 
off-diagonal components describe the Nernst-Ettingshausen effect. 
Onsager reciprocity requires that $\tensor{Q}$ in both the expressions for $\vec{j_q}$ and $\vec{E}$ are the same [see analysis following Eq.(\ref{eq:onsager})]. As an example of how our mechanism produces drag consider the Hall bar geometry, see Fig.\ref{fig1}.
When a longitudinal charge current is applied in the active layer (for $B\neq 0$) a transverse (Ettingshausen) heat current develops in both layers through efficient vertical energy transfer. 
Nernst voltage in the passive layer results in a longitudinal magnetodrag of a {\it negative sign}.

To obtain the electric field in layer 2  induced by current applied in layer 1, we first need to understand the coupling of temperature profiles $T_{1,2}(\vec r)$ in the two layers. Energy transport in the system can be described by
\bea
&& -\nabla \tensor{\kappa}_1 \nabla \delta T_1+ a(\delta T_1- \delta T_2)+\lambda \delta T_1 = - \nabla\cdot \big(\tensor{Q}^{(1)} \vec j\big)
\nonumber\\
&& -\nabla \tensor{\kappa}_2 \nabla \delta T_2+ a (\delta T_2-\delta T_1)+\lambda \delta T_2 = 0
\label{eq:etransport}
\eea
with $a$  the energy transfer rate between the two layers \cite{edrag}, $\lambda$ the electron-lattice cooling rate, and $\delta T_{i} = T_i - T_0$
(here $T_0$ is the lattice temperature, equal for  both layers). 

Here we focus on a Hall-bar geometry of two parallel rectangular layers of dimensions $L\times W$, $L\gg W$, pictured in Fig.\ref{fig1}a. For the sake of simplicity, we treat the electric and heat currents as independent of the $x$ coordinate along the bar. In layer 1, current is injected at $x=-L/2$ and drained at $x=L/2$. In layer 2, the Hall drag voltage arising across the device, $V_H$ 
and the longitudinal drag voltage, $V_\parallel$, are evaluated as
\be
V_H=\int_{-W/2}^{W/2} E_y^{(2)}dy, \quad V_\parallel= \frac{L}{W} \int_{-W/2}^{W/2} E_x^{(2)} dy
\label{eq:voltage}
\ee
The electric and thermal variables may depend on the transverse coordinate $y$, see below.

Boundary conditions for a Hall bar require 
electric current being
tangential to the side boundaries, $y=\pm W/2$, and zero temperature imbalance at the ends, $x=\pm L/2$, reflecting that the current and voltage contacts act as ideal heat sinks. The electric current parallel to the boundaries $y=\pm W/2$
gives rise to the Ettingshausen heat current that may have a component transverse to the Hall bar. The divergence of this heat current, appearing on the right hand side of Eq.(\ref{eq:etransport}), acts as an effective boundary delta function source in the heat transport equations. Boundary conditions can profoundly influence the symmetry of the resultant drag resistivity, see below.

We consider the case of a spatially uniform $\tensor Q$ in both layers. The ideal heat sinks at $x =\pm L/2$ mean that no temperature imbalance develops in the $x$-direction (except for some ``fringing'' heat currents near the contacts which give a contribution small in $W/L\ll1$, which we will ignore  in the following discussion). Since no temperature gradients are sustained in the $x$-direction for from the ends, we can reduce our problem Eq.(\ref{eq:etransport}) to a quasi-1D problem with temperature profiles that only depend on the $y$-coordinate. As a result, the only heat source arises from the Ettingshausen effect $\tensor Q^{(1)}\vec  j  = (Q^{(1)}_{yx} j)\hat {\vec y} $.

To describe transport in the presence of such a source, we will expand temperature variables in both layers in a suitable orthonormal set of functions. Here it will be convenient to use eigenstates of the operator $\p_y^2$ with zero Neumann boundary conditions at $y=\pm W/2$, given by 
\[
u_n(y)=A\cos\lp \frac{2\pi n}{W} y\rp,\quad v_n(y)=A\sin\lp \frac{2\pi (n+\frac12)}{W} y\rp
,
\]
$A=(2/W)^{1/2}$, $n=0,1,2...$ From the symmetry of the source in Eq.(\ref{eq:etransport}) we expect  $\delta T_{1,2}(y)$ to be odd in $y$. Thus only the functions  $v_n(y)$ are relevant, giving 
\[
\delta T_{1,2} (y)=\sum_{q_n}  \delta \tilde{T}_{1,2} (q_n) A\sin q_n y,\quad
q_n= \frac{2\pi (n+\frac12)}{W}.
\] 
For each $n$ we obtain a pair of algebraic equations
\bea
&& q_n^2 \kappa_1 \delta \tilde T_1 + a(\delta \tilde T_1- \delta\tilde T_2 )+\lambda \delta \tilde T_1 = F_n
\nonumber\\
&&q_n^2 \kappa_2 \delta \tilde T_2 + a(\delta \tilde T_2  - \delta \tilde T_1)+\lambda \delta \tilde T_2 = 0
\label{eq:etransportfourier}
\eea
where $\kappa_{1,2}=\kappa_{xx}^{(1,2)}$ and $F_n=2A(-1)^nQ^{(1)}_{yx} j $.
Solving Eq.(\ref{eq:etransportfourier}), we find the temperature profile in layer 2:
\be
\delta T_2 ( y) = \sum_{n\ge 0} \frac{a F_n}{L_1(q_n)L_2(q_n) -a^2}v_n(y), 
\ee
where $L_i (q_n)= \kappa_iq_n^2 + a + \lambda$ ($i=1,2$).
Since electron-lattice cooling is very slow \cite{graham2013,betz2013}, with the corresponding cooling length values in excess of few microns, we will suppress $\lambda$ in what follows. Because the boundaries in the transverse ($y$-direction) are free, a finite temperature imbalance between the edges can arise, given by $\Delta T = \delta T_2(y=W/2) - \delta T_2(y=-W/2)$. We find
\be
\Delta T = 
4A^2\sum_{n\ge 0} \frac{a Q^{(1)}_{yx} j}{L_1L_2 -a^2}
=\frac{8}{W \tilde \kappa}\sum_{n\ge 0} \frac{Q^{(1)}_{yx} j}{q_n^2(1+ \xi^2q_n^2)},
\ee
where we defined $\tilde \kappa=\kappa_1+\kappa_2$ and a length scale $\xi = \sqrt{\kappa_1\kappa_2/a\tilde \kappa}$. We evaluate the sum using the identity
$
\sum_{n=0}^{\infty}  \frac{1}{(n+\frac12)^4 + c^2 (n+\frac12)^2} = \frac{\pi^2}{2c^2} \Big( 1 - \frac{{\rm tanh} \pi c}{\pi c} \Big)
$
to obtain 
\be
\Delta T = \frac{WQ^{(1)}_{yx} j}{\tilde \kappa}  G(\xi)
,\quad
G(\xi)= 1 - \frac{2\xi}{W} {\rm tanh} \lp \frac{W}{2\xi} \rp 
.
\ee
Connecting $\Delta T$ with the drag voltage, and in particular determining its sign, requires taking full account of Onsager reciprocity. This analysis is presented below.

In the same way that the applied charge current, $\vec j$, in layer 1 causes a heat current (Peltier/Ettingshausen), a temperature imbalance in layer 2, $\Delta T$, can sustain voltage drops across the sample (Thermopower/Nernst). These two effects are related by Onsager reciprocity constraints. The cross couplings in the coupled energy and charge transport equations \cite{callen} arise from 
\be
\left( \begin{array}{cc}   -\vec j\\   \vec{j_q} \end{array}\right)  =  \left( \begin{array}{c|c} e\tensor{L}_{11}/T & e\tensor{L}_{12} \\ \hline \tensor{L}_{21}/T  & \tensor{L}_{22}  \end{array}\right)  \left( \begin{array}{cc}   \vec\nabla \mu \\  \vec \nabla \frac{1}{T}  \end{array}\right) 
\label{eq:jandjq}
\ee
where $\tensor{L}$ are $2\times2$ matrices and $e$ is the carrier charge. In this notation, the electrical conductivity equals $\tensor{\sigma} = e^2 \tensor{L}_{11}/T $, and thermal conductivity is $\tensor{\kappa} = \tensor{L}_{22}/T^2$. Comparing to the heat current due to an applied charge current, Eq.(\ref{eq:jqande}), we identify $\tensor{L}_{21} = - e\tensor{Q} \tensor{L}_{11}$.

Onsager reciprocity demands that the cross-couplings obey
$
\tensor{L}_{12}(B) = \tensor{L}_{21}^{\rm T} (-B)
$
where $B$ is the applied magnetic field (note the transposed matrix). In an isotropic system the off-diagonal components of $\tensor{L}$ obey $\tensor{L}^{(xy)} (B) =  \tensor{L}^{(yx)} (-B) $.  As a result, Onsager reciprocity reduces to 
\be 
\tensor{L}_{12}(B) = \tensor{L}_{21} (B) 
\label{eq:onsager}
\ee 
in an isotropic system. Applying Eq. \ref{eq:onsager} to Eq. \ref{eq:jandjq} in an open circuit, we find 
$
\vec E = -e^{-1}\nabla \mu = T^{-1}\tensor{L}^{-1}_{11} \tensor{Q} \tensor{L}_{11} \nabla T. 
$
For an isotropic system $\tensor{Q} = Q_{xx} \unit  + i Q_{xy} \sigma_2$,  $\tensor{L} = L_{xx} \unit  + i L_{xy} \sigma_2$, so that $[ \tensor Q, \tensor L] =0$, which gives Eq.(\ref{eq:jqande}). 

Several different regimes arise depending on the relation between the interlayer cooling length $\xi$ and the bar width $W$. Using Eq.(\ref{eq:voltage}) and  Eq.(\ref{eq:jqande}) we obtain
\be
 \left(\begin{matrix} V_\parallel \\ V_H \end{matrix} \right) = \left(\begin{matrix} R_\parallel^{\rm drag} & -R_H^{\rm drag} \\   R_H^{\rm drag} & R_\parallel^{\rm drag} \end{matrix} \right)  \left(\begin{matrix} I_\parallel \\ 0 \end{matrix} \right),
 \ee
giving the magnetodrag and Hall drag resistance values
\be
R_H^{\rm drag} = \frac{ -G(\xi)}{T \tilde \kappa} Q_{xy}^{(1)} Q_{xx}^{(2)}, \quad R_\parallel^{\rm drag} = \frac{ - LG(\xi)}{WT \tilde \kappa} Q_{xy}^{(1)} Q_{xy}^{(2)},
\label{eq:draguniform}
\ee
where we used $Q_{xx} = Q_{yy}$ and $Q_{xy} = - Q_{yx}$ for an isotropic system.  
 For a narrow bar (or, slow cooling), we have $\xi/W \gg1$ and $G\to 0$, giving vanishingly small  $R_{H,\parallel}^{\rm drag}$. For a wide bar (or, fast cooling) we have $G \to 1$ so that $R_{H,\parallel}^{\rm drag}$ saturates to a universal value independent of the interlayer cooling rate. For typical device parameters, we estimate $\xi \approx 40 \, {\rm nm}$ at $T= 300 \, {\rm K}$ \cite{edrag}. Since $L$, $W$ are a few mircons for typical graphene devices, we expect them to be firmly in the $G=1$ regime, with the Hall drag and magnetodrag attaining universal values independent of the electron-electron interaction strength.

To describe the density and $B$ field dependence, we use a simple model for $\tensor Q $.
Measurements indicate\cite{zuev, wei} that thermopower and the Nernst effect in graphene are well described by the Mott formula \cite{jonson}, giving
\be \label{eq:mott}
\tensor Q = \frac{\pi^2}{3e} k_B^2 T^2 \tensor \rho \frac{\partial [\tensor \rho^{-1}]}{\partial \mu}
, \quad 
\tensor \rho =  \left(\begin{matrix} \rho_\parallel & \rho_H \\   -\rho_H & \rho_\parallel \end{matrix} \right)
,
\ee 
with  $\rho$ the resistivity, $e<0$  the electron charge, and $\mu$ the chemical potential. 
We use a simple phenomenological model \cite{abanin} relevant for classically weak $B$ fields:
\be
\rho_\parallel= \frac{\rho_0}{\sqrt{1+ n^2/n_0^2}} , \quad \rho_H= \frac{-Bn}{e(n^2+ n_0^2)},      
\label{eq:resistivity}
\ee
where $\rho_0$ is the resistivity peak value at the Dirac point, $n$ is the carrier density, and parameter $n_0$ accounts for broadening of the Dirac point due to disorder. We account for disorder broadening of the density of states, $dn/d\mu  = (n^2+ n_0^2)^{1/4} \lp 2/ (\pi \hbar^2 v_F^2)\rp^{1/2}$.

From Eqs.(\ref{eq:draguniform}),(\ref{eq:mott}),(\ref{eq:resistivity}) and the Wiedemann-Franz relation 
for $\kappa$,
we obtain $\rho_{\parallel}^{\rm drag} = (W/L) R_{\parallel}^{\rm drag}$ and $R_{H}^{\rm drag}$  (see Fig.\ref{fig1}b,c and Fig.\ref{fig2}b,c, respectively). 
In that, we used the parameter values $n_0 = 10^{11} \, {\rm cm}^{-2}$, $\rho_0 =  \frac{h}{3e^2}$, and  a representative temperature, $T=150 \, {\rm K}$. These values match device characteristics (disorder broadening, $n_0$, and peak resistivity, $\rho_0$) described in Ref.\cite{geim}. As a sanity check, we plot the components of $\tensor Q$ (in Fig.\ref{fig2}d) which show the behavior near CN matching thermopower and Nernst effects measured in graphene\cite{zuev,wei}.

Analyzing magnetodrag, we find that $\rho_{\parallel}^{\rm drag}$ peaks at dual CN, taking on large and {\it negative} values (Fig.\ref{fig1}b,c).
Magnetodrag peak exhibits a steep $B$ dependence, $\rho_{\parallel,{\rm peak}}^{\rm drag}\propto -B^2$,  bearing a striking resemblance to measurements reproduced  in Fig.\ref{fig1}d. In particular, our model explains the negative sign of the measured magnetodrag.

Hall drag is large and sign-changing (see Fig.\ref{fig2}b,c), taking on values consistent with 
measurements\cite{Halldrag_private_communication}. Interestingly, the map in Fig.\ref{fig2}b indicates that  the sign of $R_{H}^{\rm drag}$ is controlled solely by carrier density in layer 2, breaking the $n_1\leftrightarrow n_2$
symmetry between layers. This behavior does not contradict Onsager reciprocity, it arises as a consequence of
the asymmetric boundary conditions for the Hall bar: free boundary at $y=\pm W/2$ and ideal heat sinks at the ends, $\delta T(x=\pm L/2) = 0$. This allows for finite temperature gradients to be sustained across the bar but not along the bar, see Fig.\ref{fig2}a.

For other geometries, the temperature gradient can be obtained by balancing the heat flux due to thermal conductivity against the net heat flux in the two layers, 
$(\kappa_1+\kappa_2)\nabla \delta T= \tensor D \, \tensor{Q}^{(1)} \vec j_1$
(see Eq.(24) of Ref.\cite{magneto-dima}). The quantity $\tensor D$
can in principle be obtained by solving heat transport equations. 
Adopting the same approach as above, we find a magneto and Hall-drag resistivity 
\be
\tensor{\rho}^{\rm drag} = \frac{1}{T \tilde{\kappa} }\tensor{Q}^{(2)} \tensor D \, \tensor Q^{(1)},  \quad \vec E_2 = \tensor{\rho}^{\rm drag} \vec j_1.
\label{eq:dragrho}
\ee
For isotropic heat flow, $\tensor D = \unit$. In this case,  since 
$\tensor Q^{(1)}$ and $\tensor Q^{(2)}$ commute, the resulting drag  is layer-symmetric, $n_1 \leftrightarrow n_2$\cite{magneto-dima}. In particular, Hall drag for $\tensor D = \unit$  vanishes on the diagonal $n_1 = -n_2$. In contrast, for anisotropic heat flow, such as that discussed above,
we expect a generic tensor $\tensor D \ne \unit$ and thus no layer symmetry.

We wish to clarify, in connection to recent measurements,\cite{Halldrag_private_communication} that 
layer symmetry
$n_1 \leftrightarrow n_2$
 implies a swap of current and voltage contacts. 
Layer symmetry, which implies $\tensor D = \unit$ in Eq.(\ref{eq:dragrho}), will
therefore 
only hold for Hall bars equipped with wide voltage contacts, for which
the contact and the bar widths are comparable. This is indeed the case for
the cross-shaped devices used in Ref.\cite{geim}. However it is not the case for
a Hall bar with {\it noninvasive} voltage probes which are much narrower
than the bar width, as assumed in our analysis above. Noninvasive probes, which  have little effect on temperature distribution in the electron system, translate into $\tensor D \ne \unit$ and no layer symmetry.

In summary, magnetic field has 
 dramatic effect on drag at CN
because it induces  strong 
coupling between neutral and charge modes,
which are completely decoupled in the absence of 
magnetic field 
in a uniform system. Field-induced mode coupling leads to giant drag
that dwarfs the conventional momentum drag contribution
as well as a remnant drag due to spatial inhomogeneity\cite{edrag}.
Our estimates indicate that these two contributions are orders of magnitude
smaller than the predicted magnetodrag, which also has an opposite sign.
The giant magnetodrag and Hall drag values attained at classically weak
magnetic fields, along with the unique density dependence and sign, make
these effects easy to identify in experiment. The predicted magnetodrag is in good agreement with findings in Ref.\cite{geim}. Magnetic field, coupled with drag
measurements at CN, provides a unique tool for probing the neutral modes in
graphene.

We acknowledge useful discussions with A. K. Geim, P. Jarillo-Herrero, L. A. Ponomarenko, and financial support from the NSS program, Singapore (JS).

\end{document}